\input{aipcheck}

\documentclass[
    ,final            
  ]
  {aipproc}

\layoutstyle{6x9}

\begin{document}

\def\lsim{\lower.5ex\hbox{$\; \buildrel < \over \sim \;$}}
\def\gsim{\lower.5ex\hbox{$\; \buildrel > \over \sim \;$}}

\title{Predicting spin of compact objects from their QPOs: A global QPO model}


\author{Banibrata Mukhopadhyay}{
  address={Astronomy and Astrophysics Programme, 
Department of Physics, Indian Institute of Science, Bangalore-560012\\
bm@physics.iisc.ernet.in}
}


\begin{abstract}
We establish a unified model to explain Quasi-Periodic-Oscillation (QPO)
observed from black hole and neutron star systems globally. This is based
on the accreting systems thought to be damped harmonic oscillators 
with higher order nonlinearity. The model explains multiple
properties parallelly independent of the nature of the compact object. 
It describes QPOs successfully 
for several compact sources. Based on it, we predict the spin frequency of the 
neutron star Sco~X-1 and the specific angular momentum of black holes GRO~J1655-40, 
GRS~1915+105.
\end{abstract}

\maketitle


\section{Introduction}

The origin of Quasi-Periodic-Oscillation (QPO) and its properties are still ill-understood. 
The observed QPO frequencies of compact objects
are expected to be related to the spin parameter of the compact object itself.
In certain neutron star systems a pair of QPO forms and QPO frequencies appear
to be separated either by the order of the spin frequency of the neutron star,
apparently for {\it slow rotators} e.g. 4U~1702-429, or by
half of the spin frequency, for {\it fast rotators} e.g. 4U~1636-53.
Their frequency separation decreases
with the increase of one of the QPO frequencies. The black hole systems, on the
other hand, e.g. GRO~J1655-40, XTE~J1550-564, GRS~1915+105, 
exhibit the kHz QPO pairs which seem to appear at a $3:2$ ratio
\cite{rem02}. 

Interaction between the surface current density in the disk and the stellar magnetic field
generating warps and subsequent precession instability in the
inner accretion disk produces motions at low frequencies.
This is similar to the Lense-Thirring precession \cite{sv98}.
This can explain the mHz QPOs
for strongly magnetized neutron stars \cite{sl02}.
The effect of nonlinear coupling between
g-mode oscillations in a thin relativistic disk and warp has also been examined \cite{kat04}
for a static compact object.
Recent observations \cite{pal-mau} indicate a strong correlation between low and high 
frequency QPOs holding over mHz to kHz range
which strongly supports the idea that QPOs are universal physical processes independent of the
nature of the compact object.
The correlation is also explained
in terms of centrifugal barrier model \cite{tw02}.
Indeed earlier the correlation was shown in terms of the effective boundary wall 
created by the strong centrifugal force in the disk \cite{skc-works}.

Accretion dynamics is a nonlinear hydrodynamic/magnetohydrodynamic phenomenon.
It was already shown that QPOs may arise from nonlinear resonance phenomena
in an accretion disk (e.g. \cite{kluz04,pet05})
occurred due to resonance between epicyclic motions of accreting matter.
However, the separation between vertical and radial epicyclic frequencies increases 
mostly as a function of either of the epicyclic frequencies
contradicting the observed QPO feature.

We aim at establishing a {\it global} model based on
higher order nonlinear resonance theory
to describe black hole and neutron star QPOs together and 
then try to predict the spin
parameter/frequency of compact objects. 
The model predicts, alongwith important QPO features, the spin parameter of black holes 
as well as spin frequency of a neutron star.

\section{Model}

A system of $N$ degrees of freedom has $N$ linear natural frequencies $\nu_i$; $i=1,2,...N$ \cite{nm79}.
These frequencies have commensurable relations 
which may cause the corresponding modes to be strongly coupled and
yield an internal resonance. If the system is excited by an external frequency 
$\nu_*$, then the commensurable relation exhibiting resonance might be
\begin{eqnarray}
a\nu_*=\sum_{i=1}^N b_i \nu_i,\,\, {\rm with}\,\,\, a+\sum_{i=1}^N|b_i|=j,
\label{frqrel}
\end{eqnarray}
apart from all the primary and secondary resonance conditions $c\nu_*=d\nu_m$,
where $a,b_i,c,d$ are integers and
$j=k+1$ when $k$ is the order of nonlinearity.

Now we consider an accretion disk with a possible higher order nonlinear resonance. 
The resonance is driven by the combination of the strong disturbance created by the rotation
of the compact object with spin $\nu_s$ and the existent (weaker) disturbances in
the disk at the frequency
of the radial ($\nu_r$) and vertical ($\nu_z$)
epicyclic oscillations given by (e.g. \cite{st83})
\begin{eqnarray}
\nu_r=\frac{\nu_o}{r}\sqrt{\Delta-4(\sqrt{r}-a)^2},~
\nu_z=\frac{\nu_o}{r}\sqrt{r^2-4a\sqrt{r}+3a^2},
\label{epi}
\end{eqnarray}
where $\Delta=r^2-2Mr+a^2$, $a$ is the specific angular momentum (spin parameter)
of the compact object and $\nu_o$ is the orbital frequency of the disk particle given by
$2\pi\nu_o=\Omega=1/(r^{3/2}+a)$.

\begin{figure}
  \includegraphics[height=.3\textheight,width=0.5\textwidth]{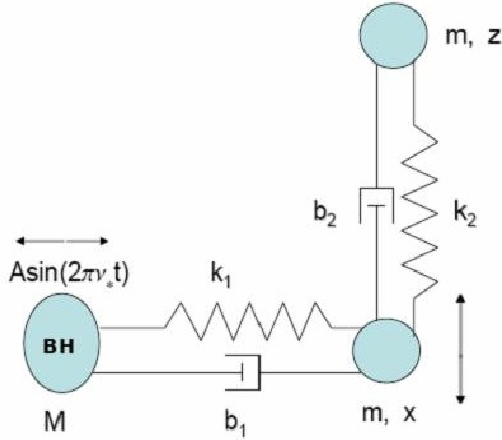}
  \caption{
Cartoon diagram describing coupling of various modes
in an accretion disk and corresponding nonlinear oscillator. The oscillators
describing by spring constant $k_1$ and $k_2$ indicate
respectively the coupling of spin frequency $\nu_*$ of the compact object with mass $M$
to radial ($x$) epicyclic frequency
and vertical ($z$) epicyclic frequency of a disk blob with mass $m$,
where $b_1$ and $b_2$ represent corresponding damping factors respectively.
\label{figcart}
}
\end{figure}

We describe the system schematically in Fig. \ref{figcart}. It is composed of
two oscillators due to radial and vertical epicyclic motion
with different spring constants.
The basic idea is that the mode corresponding to $\nu_s$ 
will couple to the ones corresponding to $\nu_r$ and $\nu_z$
igniting new modes with frequencies $\nu_{r,z}\pm p\nu_s$ \cite{nm79},
where $p$ is a number e.g. $L/2$ or $L$, when $L$ is an integer, 
if the effect is nonlinear or linear \cite{nm79,kluz04} respectively.
While $L=1$ corresponds to the dominant interaction, modes with other $L$ are very weak to exhibit
any observable effect.
Now at a certain radius in the nonlinear regime where
$\nu_s/2$ (or $\nu_s$ in the linear regime) is close to $\nu_z-\nu_r$ \cite{nm79,kluz04},
and $\nu_s/2$ (or $\nu_s$) is also coincidentally close to the frequency difference
of any two newly excited modes, a resonance may occur which locks the frequency difference
of two excited modes at $\nu_s/2$ (or $\nu_s$).

As the neutron star has a magnetosphere coupled with the accretion disk, the mode with $\nu_s$ can
easily disturb the disk matter. 
A black hole, on the other hand, with magnetic field
connecting to the surrounding disk, may transfer energy and angular momentum 
to the disk by the variant Blandford-Znajek process \cite{bz77}
what was already verified by XMM-Newton observation (e.g. \cite{wil-li}).
This confirms possible ignition of new modes in a disk around
a spinning a black hole. 

Therefore, rewriting (\ref{frqrel}) for an accretion disk we obtain
\begin{eqnarray}
\nu_r+\frac{n}{2}\nu_s=\nu_z-\frac{\nu_s}{2}+\frac{m}{2}\nu_s
\label{frqrelac}
\end{eqnarray}
with $-b_1=b_2=2$ and $m,n$ are integers. Now we propose the higher and lower QPO frequency
of a pair respectively to be
\begin{eqnarray}
\nu_h=\nu_r+\frac{n}{2}\nu_s,\,\ \nu_l=\nu_z-\frac{\nu_s}{2}.
\label{qpofrq}
\end{eqnarray}
Hence, we understand from (\ref{frqrel}) and (\ref{frqrelac})
the order of nonlinearity
in accretion disks exhibiting QPOs $k=n-m+4$.

At an appropriate radius where $\nu_z-\nu_r\sim\nu_s/2$ and resonance
is supposed to take place, $\Delta\nu=\nu_h-\nu_l\sim\nu_s/2$ for $n=m=1$, and 
$\Delta\nu\sim\nu_s$ for $n=m=2$.
$n=1$ implies a nonlinear coupling between the radial
epicyclic mode and the disturbance due to spin of the neutron star, which results in
$\Delta \nu$ locking in the nonlinear regime with $m=1$. On the other hand, $n=2$
implies a linear coupling resulting $\Delta \nu$ locking in the linear regime with $m=2$.
For a black hole, however, in absence of
its magnetosphere, disturbance and then corresponding coupling may not be
strongly nonlinear and occurs with the condition $\nu_z-\nu_r\lsim\nu_s$ resulting
the resonance locking at the linear regime
with $n=m=2$, which produces $\Delta \nu\lsim\nu_s$ (sometimes $\sim2\nu_s/3$).
However, if we enforce the resonance strictly to occur at marginally stable circular orbit,
then it occurs in the nonlinear regime with
$\Delta \nu\lsim\nu_s/2$ (sometime $\sim\nu_s/5$) and $\nu_z-\nu_r\lsim\nu_s$
for $n=m=1$. 
In principle, there may be possible
resonances with other combination of $n$ and $m$ (e.g. $n=2, m=1$) which are expected to be
weak to observe.

Once we know the spin frequency $\nu_s$ of a neutron star from observed data 
we can determine specific angular momentum $a$ (spin parameter)
with the information of equatorial radius $R$, spin frequency $\nu_s$, mass $M$, 
radius of gyration $R_G$.
If we consider the neutron star to be spherical in shape with equatorial
radius $R$, then
the moment of inertia and spin parameter are computed as
$I=MR_G^2,\,\,\,a=I\Omega_s c/GM^2$,
where $\Omega_s=2\pi\nu_s$, $G$ is the Newton's gravitation constant and $c$ is speed of light.
We know that for a solid sphere $R_G^2=2\,R^2/5$ and for a hollow sphere $R_G^2=2\,R^2/3$. 
However, in practice for a neutron star $R_G^2$ should be in between.
Moreover, the shape of a very fast rotating
neutron star is expected to be deviated from spherical to ellipsoidal.
Hence, in our calculation we restrict $0.35\le (R_G/R)^2 \le0.5$.

On the other hand, for a black hole QPO
$a$ is the most natural quantity what we supply as an input.
The corresponding angular
frequency of a test particle at 
the radius of marginally stable circular orbit $r_{ms}$ in spacetime around it
is then given by \cite{st83}
\begin{eqnarray}
\Omega_{BH}=2\pi\nu_s=-\frac{g_{\phi t}(r=r_{ms})}{g_{\phi\phi}(r=r_{ms})}=
\frac{2a}{r_{ms}^3+r_{ms}\,a^2+2a^2},
\label{bhfreq}
\end{eqnarray}
where and light inside $r_{ms}$ is practically
not expected to reach us. 

\section{Results}
\subsection{Neutron stars}

The choice of mass $M=1.4M_\odot$ and $(R_G/R)^2=0.4$
does not suffice the observation mostly. Hence,
we consider other values of parameters given in TABLE 1.

For 4U~1636-53, 
our theory has an excellent agreement with observed data
in accordance with realistic EOS \cite{eos} shown in Fig. \ref{fig2}a and given in TABLE 1.
On the other hand, for 4U~1702-429, 
if the star is considered to be an ellipsoid and/or not to be a solid sphere and/or 
to have mass $M< 1.4M_\odot$, then our theory has an excellent agreement with observed data
with realistic $R$ \cite{eos} shown in Fig. \ref{fig2}b.
Similarly for other neutron
stars showing twin kHz QPOs, results from our model have good agreement with observed data
(not discussed in the present paper but would be described elsewhere in future), which
are the beyond scope to discuss in the present paper.

\subsubsection{Estimating spin of Sco~X-1}

\begin{figure}
  \includegraphics[height=.3\textheight,width=0.8\textwidth]{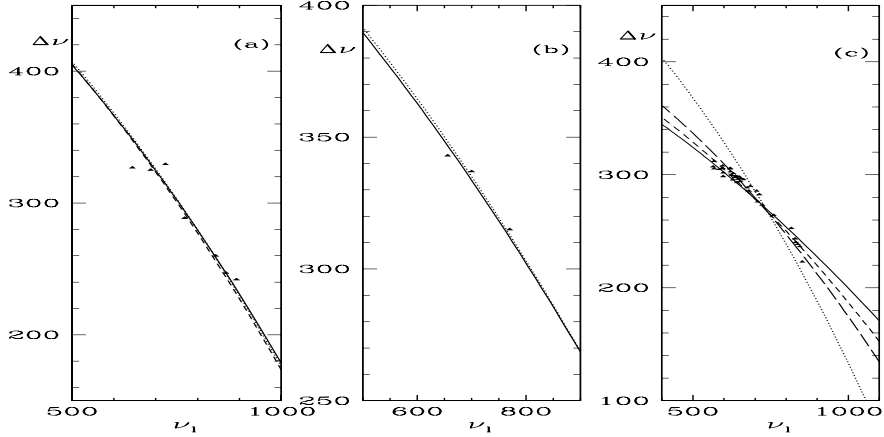}
\caption{Variation of QPO frequency difference in a pair as a function of lower QPO frequency for
(a) 4U~1636-53, (b) 4U~1702-429, (c) Sco~X-1. Results for parameter sets given in TABLE 1 
from top to bottom row for a particular neutron star correspond to solid, dotted, dashed (4U~1702-429
and Sco~X-1), long-dashed (Sco~X-1) lines.
The triangles are observed data points.
}
\label{fig2}
\end{figure}

The spin frequency of Sco~X-1 is not known yet.
We compare in Fig. \ref{fig2}c the observed variation of frequency separation as a 
function of lower QPO frequency with that obtained from our model.
We find that mass of Sco~X-1 must be less than $1.4M_\odot$
and results with sets of inputs with smaller $\nu_s$ and $M$
fit the observed data better and argue that Sco~X-1 is a slow rotator with $\nu_s\sim 280-300$.

\begin{table}
\begin{tabular}{ccccccccccccccccccccccccccccccccccccccc}
\hline
neutron star & $\nu_s$ & $M$ & $(R_G/R)^2$ & $n,m$ & $R$ & range of $r_{QPO}$ \\
\hline
4U~1636-53 & $581.75$ & $1.4$ & $0.5$ & $1$ & $16.5$ & $6.7-7.7$\\
4U~1636-53 & $581.75$ & $1.1$ & $0.4$ & $1$ & $14.3$ & $8.2-9.3$ \\
4U~1636-53 & $581.75$ & $1.2$ & $0.35$ & $1$ & $16.8$ & $7.7-8.7$ \\
\hline
4U~1702-429 & $330.55$ & $1.0$ & $0.5$ & $2$ & $18.8$ & $10.2-11$ \\
4U~1702-429 & $330.55$ & $0.83$ & $0.35$ & $2$ & $18.5$ & $11.7-12.5$ \\
\hline
 & estimated $\nu_s$ &  &  &  & &  \\
\hline
Sco~X-1 & $422.0$ & $0.9$ & $0.5$ & $1$ & $17.4$ & $9.4-12.5$\\
Sco~X-1 & $540.0$ & $1.2$ & $0.4$ & $1$ & $16.0$ & $8-9.1$\\
Sco~X-1 & $280.0$ & $0.8$ & $0.5$ & $2$ & $17.5$ & $11.1-14.4$ \\
Sco~X-1 & $292.0$ & $0.81$ & $0.35$ & $2$ & $18.6$ & $11.2-14$ \\
\hline
\end{tabular}
\caption{$\nu_s$ is given in unit of Hz, $M$ in $M_\odot$, $R$ in km,
radial coordinate where QPO occurs, $r_{QPO}$, in unit of Schwarzschild radius.}
\label{tab1}
\end{table}

\subsection{Black holes}

The possible mass or range of mass of several black holes
is already predicted from observed data.
However, the spin of them is still not well established.
By supplying predicted mass and arbitrary values of $a$
our theory reproduces observed QPOs for GRO~J1655-40 and GRS~1915+105 with their $3:2$ ratio 
for $n=m=2$ at a radius outside $r_{ms}$ given in TABLE 2. 
However, if we enforce QPOs
to produce at $r_{ms}$ strictly, then they produce at a higher $a$ for $n=m=1$. 
Similarly, results from our model for other black holes showing twin kHz QPOs have good agreement 
with observed data (would be discussed elsewhere in future), which
are the beyond scope to discuss in the present paper.

\begin{table}
\hspace{-2cm}
\begin{tabular}{cccccccccccccc}
\hline
{\small black hole $M$} & $a$ & $\nu_h$ & $\nu_l$
& $r_{QPO}$ & $\Delta r$ & $n,m$ \\
\hline
{\small GRO~J1655-40} & estimated & theory/observation & theory/observation & & &\\
\hline
$6-7$ & $0.737-0.778$ & $450/450$ & $300/300$ & $4.93-4.25$ & $1.71-1.23$ & $2$\\
$7.05$ & $0.95$ & $451.31/450$ & $299.04/300$ & $1.94$ & $0$ & $1$\\
\hline
{\small GRS~1915+105} & estimated & theory/observation & theory/observation & & &\\
\hline
$10-20$ & $0.606-0.797$ & $168/168$ & $113/113$ & $7.38-3.9$ & $3.58-0.98$ & $2$\\
$18.4$ & $0.95$ & $167.35/168$ & $114.61/113$ & $1.94$ & $0$ & $1$\\
\hline
\end{tabular}
\caption{$\nu_{l,h}$ are given in unit of Hz, $M$ in $M_\odot$,
$r_{QPO}$ and its distance from marginally stable orbit, $\Delta r$,
are expressed in unit of Schwarzschild radius.
}
\label{tab2}
\end{table}


\section{Summary}

We have prescribed a global QPO model based on nonlinear resonance mechanism in accretion disks.
Based on this we have predicted the spin parameter/frequency 
of compact objects. 
The model has addressed, for the first time to best of our knowledge, the variation of QPO frequency 
separation in a pair as a function of the QPO frequency itself for neutron stars successfully. 
We argue that Sco~X-1 is a slow rotator.

We have addressed QPOs of black holes as well
and predict their spin parameter ($a$) which is not
well established yet.
According to the present model, none of them is an extremally rotating black hole.
As our model explains QPOs observed from several black holes and neutron stars 
including their specific properties,
it favors the idea of QPOs to originate
from a unique mechanism, independent of the nature of compact objects.

\begin{theacknowledgments}
The work was supported partly by a project, Grant No. SR/S2/HEP12/2007, funded by DST, India.
\end{theacknowledgments}



\bibliographystyle{aipproc}   

\bibliography{sample}

\IfFileExists{\jobname.bbl}{}
 {\typeout{}
  \typeout{******************************************}
  \typeout{** Please run "bibtex \jobname" to optain}
  \typeout{** the bibliography and then re-run LaTeX}
  \typeout{** twice to fix the references!}
  \typeout{******************************************}
  \typeout{}
 }



\end{document}